\documentclass[article,reprint,superscriptaddress]{revtex4-1}
\usepackage{graphicx}
\usepackage{longtable}
\usepackage{epstopdf}
\usepackage{dcolumn}
\newcolumntype{x}{D{/}{/}{4.4}}
\newcolumntype{y}{D{.}{.}{4.13}}
\newcolumntype{z}{D{.}{.}{4.14}}
\newcolumntype{e}{D{.}{.}{8.10}}
\newcolumntype{f}{D{.}{.}{5.8}}

\begin{document}

\title{Spins and Magnetic Moments of $^{49}$K and $^{51}$K: establishing the 1/2$^+$ and 3/2$^+$ level ordering beyond $N$ = 28.}

\author{J. Papuga}
\affiliation{Instituut voor Kern- en Stralingsfysica, KU Leuven, B-3001 Leuven, Belgium}
\author{M. L. Bissell}
\affiliation{Instituut voor Kern- en Stralingsfysica, KU Leuven, B-3001 Leuven, Belgium}
\author{K. Kreim}
\affiliation{Max-Plank-Institut f\"{u}r Kernphysik, D-69117 Heidelberg, Germany}
\author{K. Blaum}
\affiliation{Max-Plank-Institut f\"{u}r Kernphysik, D-69117 Heidelberg, Germany}
\author{B.A. Brown}
\affiliation{Department of Physics and Astronomy and National Superconducting Cyclotron Laboratory, Michigan State University,
East Lansing, Michigan 48824, USA}
\author{M. De Rydt}
\affiliation{Instituut voor Kern- en Stralingsfysica, KU Leuven, B-3001 Leuven, Belgium}
\author{R. F. Garcia Ruiz}
\affiliation{Instituut voor Kern- en Stralingsfysica, KU Leuven, B-3001 Leuven, Belgium}
\author{H. Heylen}
\affiliation{Instituut voor Kern- en Stralingsfysica, KU Leuven, B-3001 Leuven, Belgium}
\author{M. Kowalska}
\affiliation{Physics Department, CERN, CH-1211 Geneva 23, Switzerland}

\author{R. Neugart}
\affiliation{Institut f\"{u}r Kernchemie, Universit\"{a}t Mainz, D-55128 Mainz, Germany}
\author{G. Neyens}
\affiliation{Instituut voor Kern- en Stralingsfysica, KU Leuven, B-3001 Leuven, Belgium}
\author{W. N\"{o}rtersh\"{a}user}
\affiliation{Institut f\"{u}r Kernchemie, Universit\"{a}t Mainz, D-55128 Mainz, Germany}
\affiliation{Institut f\"{u}r Kernphysik, TU Darmstadt, D-64289 Darmstadt, Germany}
\author{T. Otsuka}
\affiliation{Department of Physics and Center for Nuclear Study,
University of Tokyo, Hongo, Tokyo 113-0033, Japan
and RIKEN, Hirosawa, Wako-shi, Saitama 351-0198, Japan}
\author{M. M. Rajabali}
\affiliation{Instituut voor Kern- en Stralingsfysica, KU Leuven, B-3001 Leuven, Belgium}
\author{R. S\'{a}nchez}
\affiliation{GSI Helmholtzzentrum f\"{u}r Schwerionenforschung, D-64291 Darmstadt, Germany}
\author{Y. Utsuno}
\affiliation{Japan Atomic Energy Agency, Tokai-mura, Ibaraki 319-1195, Japan}
\author{D. T. Yordanov}
\affiliation{Max-Plank-Institut f\"{u}r Kernphysik, D-69117 Heidelberg, Germany}


\date{\today}

\begin{abstract}
The ground-state spins and magnetic moments of $^{49,51}$K have been measured using bunched-beam high-resolution collinear laser spectroscopy at ISOLDE-CERN. For $^{49}$K a ground-state spin  $I = 1/2$ was firmly established.  The observed hyperfine structure of $^{51}$K requires a spin $I > 1/2$ and from its magnetic moment $\mu(^{51}\text{K})= +0.5129(22)\, \mu_N$ a spin/parity $I^\pi=3/2^+$ with a dominant $\pi 1d_{3/2}^{-1}$ hole configuration was deduced.  This establishes for the first time the re-inversion of the single-particle levels and illustrates the prominent role of the residual monopole interaction for single-particle levels and shell evolution.

\end{abstract}


\maketitle
The nuclear shell-model forms the basis for our understanding of atomic nuclei and since the very beginning spins and magnetic moments of ground states have played a crucial role \cite{Mayer48,Mayer49}. Single-particle as well as collective degrees of freedom in atomic nuclei can be described with modern large-scale shell-models \cite{Caurier2005}. However, the interplay between theory and experiment is indispensable for further improving the shell-model effective interactions as new regions of the nuclear chart are being explored.

Since more and more rare-isotope beams became available, strong modifications to the well-known shell structure were required in several regions of the nuclear chart.  Examples are the unexpected level ordering in the "island of inversion" isotopes $^{31,33}$Mg \cite{Neyens2011}, the weakening of the $N=28$ shell gap below Ca \cite{Bastin2007}, the monopole migration of proton single-particle levels towards $N=28$ \cite{Gade2006} and $N=50$ \cite{Flanagan2009}.  The origin of the changes in shell structure has been discussed in several theoretical papers \cite{Otsuka2001,Otsuka2005,Otsuka2006,Smirnova2010,Otsuka2010}. As experimental evidence is growing, effective shell-model interactions as well as mean-field models are being modified to account for the new observations \cite{Grasso2007,Nowacki2009,Zalewski2009,Kaneko2011,Wang2011}.

In the region below Ca ($Z=20)$, where protons occupy the $sd$ shell and neutrons occupy the $pf$ shell, the gradual filling of the $\nu 1f_{7/2}$ orbit from $N = 20$ to $N = 28$ leads to a strong decrease of the first excited $1/2^+$ energy in the $_{19}$K and $_{17}$Cl isotopes. This level becomes the ground state in $^{47}$K \cite{Touchard1982} and in the Cl chain the inversion begins at $^{41}$Cl \cite{Ollier2003}.  Following the explanation given by Otsuka \textit{et al.} \cite{Otsuka2005}, occupation of the $\nu 1f_{7/2}$ orbit (having $j=l+1/2$) leads to an increased binding of the $\pi 1d_{3/2}$ orbital (with $j^{'}=l^{'}-1/2$), and thus to a near-degeneracy (even inversion) with the $\pi 2s_{1/2}$ orbit as the $\nu 1f_{7/2}$ orbit is fully occupied at $N=28$.

This raises the question as to what happens beyond $N=28$, when the higher $pf$ orbits are being filled. Using a Woods-Saxon potential e.g. the monotonic lowering of the $\pi 1d_{3/2}$ energy with respect to the $\pi 2s_{1/2}$ energy is predicted to continue well beyond $N=28$.  However, it is by now also well-known that the residual monopole interaction, which changes with isospin and thus with neutron number, plays an important role in the evolution of single-particle levels.  This residual monopole interaction has a central, spin-orbit and tensor term, all of which need to be determined from experimental data.  Furthermore, these monopole interactions are strongest between orbits with the same number of nodes, and thus in the region beyond $N=28$ it is the $\pi 2s_{1/2} - \nu 2p_{3/2}$ interaction strength that is the dominant one \cite{Otsuka2010}. Until now, very little experimental data are available which probe this part of the nucleon-nucleon residual interaction.  A rather extended level scheme is available for $^{49}$K \cite{Broda2010} but all spins are tentative as long as the ground-state spin is not established. These recent in-beam data favor a ground-state spin $I=1/2$ while earlier $\beta$-decay work suggests $I=3/2$ \cite{Carraz1982}.  For $^{51}$K no excited states are known. From $\beta$-decay studies the ground-state spin is tentatively assigned to be $I=3/2$ with a dominant $\pi 1d_{3/2}$ hole structure \cite{Perrot2006}.

This letter presents the measured ground-state spins and magnetic moments of $^{49,51}$K with  two and four neutrons beyond $N=28$ respectively, thus gradually filling the $\nu 2p_{3/2}$ orbit. The K$^{+}$ beams were produced at the ISOLDE facility at CERN where 1.4 GeV protons impinged on a thick $\text {UC}_{x}$ target. Atoms diffused out of the target and were surface ionized in a heated tube at $\sim$2000 $^{\circ}$C. The ions were accelerated up to 40 kV, mass separated  and bunched by the gas-filled Paul trap (ISCOOL) \cite{Franberg2008,Flanagan2009}. Collinear laser spectroscopy \cite{Mueller1983} was performed using the 769.9 nm $4s\;{^2S_{1/2}} \rightarrow 4p\;{^2P_{1/2}}$ atomic transition in K, after the beam was neutralized via collisions with neutral K atoms in a charge exchange cell (CEC).  To scan the hyperfine structure the Doppler-shifted laser frequency, as experienced by the fast atoms, was modified by applying a tunable scanning voltage of $\pm 500$ V on the CEC. The fluorescence light emitted from the resonantly laser-excited atoms was subsequently detected using photomultiplier tubes (PMTs). By gating the PMT signals such that photons were only collected  during the $\sim$6 $\mu$s period when the  bunch of atoms pass in front of the PMT, the background photon count rate was reduced by a factor of 15000 (= ISCOOL accumulation time/ion bunch temporal width) compared to a continuous beam detection.


The relatively slow decay rate of the atomic transition ($3.8 \times 10^7$ s$^{-1}$),  low quantum efficiency (2.5\%) and high heat-related dark counts of PMTs operating at these wavelengths hinder optical spectroscopy of exotic K isotopes. In order to perform the measurements on a $^{51}$K beam of $\sim$4000 ions/s, a new optical detection station was developed. Details on this will be presented in a forthcoming paper \cite{Kreim2013}.

\begin{figure*}[t]
\includegraphics[width=0.8\textwidth]{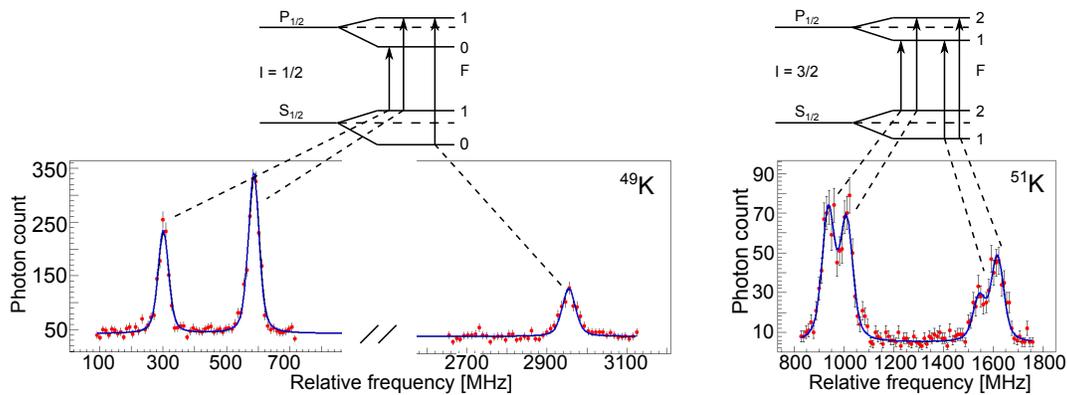}
\caption{\label{HFS-49-51K} (Color online) Typical hyperfine spectra for $^{49}$K and $^{51}$K. Spectra are shown relative to the centroid of $^{39}$K.}
\end{figure*}

Typical hyperfine structure (hfs) spectra for $^{49}$K and $^{51}$K are shown in Fig.~\ref{HFS-49-51K}. To convert the recorded scanning voltage into frequency, AME 2012 \cite{Wang2012}  masses were used with the recently measured masses of $^{49}$K \cite{Lapierre2012} and $^{51}$K \cite{Gallant2012} already included. The hfs spectra have been fitted using the $\chi^2$ minimization procedure MINUIT. The peak positions were determined by the hyperfine parameters $A(^{2}S_{1/2})$ and $A(^{2}P_{1/2})$, the nuclear spin $I$ and the center of gravity of the structure defining the isotope shift. The $A$-parameters are directly related to the nuclear magnetic moment through the relation $ A = {\mu B_{0}}/{I J}$, where $B_{0}$ is the magnetic field generated by electrons at the site of a nucleus.  Voigt profiles with common line widths have been used for all peaks in the spectrum. The nuclear spin determines the number of allowed transitions.  For $I=1/2$ only three transitions are allowed (left of Fig.~\ref{HFS-49-51K}) and the ground-state spin of $^{49}$K can be unambiguously assigned as $I=1/2$. For all other spins four resonances appear in the spectrum, as illustrated in the right part of Fig.~\ref{HFS-49-51K}. We have therefore fitted the $^{51}$K spectra assuming a spin 3/2, 5/2 and 7/2. In the case of a $J=1/2$ to $J=1/2$ transition the ratio of the upper-to-lower hyperfine parameter cannot be used to  exclude a particular spin (method used e.g. to assign the $^{72,74}$Cu spins \cite{Flanagan2010}), as the \textit{relative} hyperfine structure separations are unaffected by the nuclear spin in this case. There is, however, an indirect way to determine the $^{51}$K spin, namely by comparing the intensities of  the different hyperfine structure components. From the intensity ratios observed for  $^{51}$K  we can exclude a spin 5/2 (or even higher) at a confidence level of 95$\%$. To further confirm the measured spin $I = 3/2$, the extracted magnetic moment for different spin assumptions can be compared to single-particle moments and predictions by shell-model calculations.

\begin{figure}[t]
\includegraphics[width=0.4\textwidth]{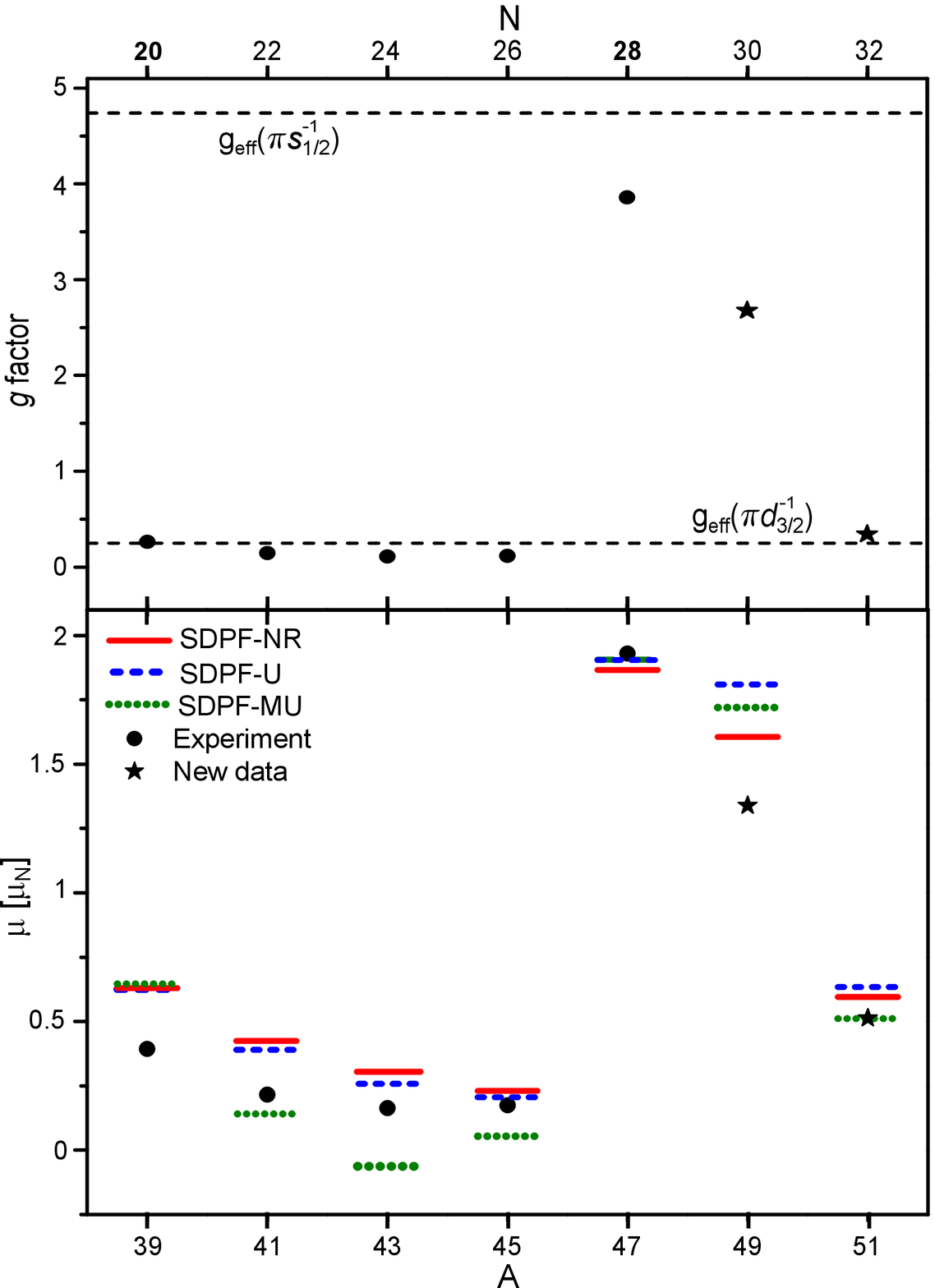}
\caption{\label{g-mu} (Color online) Experimental $g$ factors (upper) and magnetic moments (lower) from \cite{Beckmann1974,Touchard1982} (dots) and the new data (stars) for $^{49}$K (1/2) and $^{51}$K (3/2). Effective single-particle $g$ factors are calculated with $g_{s}^{\rm{eff}}$=0.85$g_{s}^{\rm{free}}$ and $g_l^\pi$=1.15, $g_l^\nu$=-0.15. See text for details on calculations.}
\end{figure}




\begin{table}[h]
\caption[]{Fitted hyperfine parameters for the studied isotopes (assuming different spins for $^{51}$K).\label{data}}
\begin{ruledtabular}
\begin{tabular}{c  x  e  f}

Isotope   
    &\multicolumn{1}{c}{$I$}  
          &\multicolumn{1}{c}{$A(^{2}S_{1/2})$ (MHz) }
              &\multicolumn{1}{c} {$A(^{2}P_{1/2})$ (MHz)}\\
\hline\\[-1em]
$^{39}$K & 3/2    & +231.0\,~\,(3) & +27.8\,(2) \\[0.1em]
$^{47}$K & 1/2    & +3413.2\,~\,(2) & +411.8\,(2) \\[0.1em]
$^{49}$K & 1/2    & +2368.2\,(14) & +285.6\,(7) \\[0.1em]
$^{51}$K & 3/2  & +302.5\,(13) & +36.6\,(9) \\[0.1em]
$^{51}$K & (5/2)  & +201.6\,~\,(9) & +24.4\,(6) \\[0.1em]
$^{51}$K & (7/2)  & +151.3\,~\,(7) &  +18.3\,(4) \\

\end{tabular} \\
\end{ruledtabular}
\end{table}

The magnetic moments are deduced from the $A(^2S_{1/2})$ values relative to that of $^{39}$K, for which a very precise measurement of $A_{\rm{ref}}(^{2}S_{1/2})$ = 230.8598601(7) MHz and $\mu_{\rm{{ref}}} = +0.3914662(3)\, \mu_{N} $ is available from an atomic beam magnetic resonance measurement \cite{Beckmann1974}. The results are presented in Table \ref{moments}. The statistical errors are smaller than the hyperfine anomalies measured for $^{40,41,42}$K relative to $^{39}$K \cite{Beckmann1974,Eisinger1952,Chan1969}. Following the approach of Bohr \cite{Bohr1951} these hyperfine anomalies are well reproduced and those for $^{47,49,51}$K are predicted to be less than $0.3\%$. This is included as an additional error on the magnetic moments (in square brackets).



\begin{table*}
\caption[]{Experimental magnetic moments (in units of $\mu_N$) and $g$ factors, compared to shell-model values using different effective interactions (see text for details). The error in square brackets represents the uncertainty related to the hyperfine anomaly. The $^{47}$K value is in good agreement with the literature value 1.933(9)\,$\mu_{N}$ \cite{Touchard1982}. Energies for the predicted states of $^{51}$K are shown in keV. \label{moments}}
\begin{ruledtabular}
\begin{tabular}{c  c  y  y  z  z  z}
Isotope 
  & $I^{\pi}$
      & \multicolumn{1}{c}{ $g_{\text {exp}}$  } 
           &\multicolumn{1}{c}{$\mu_{\text {exp}}$ }   
               &\multicolumn{1}{c}{$\mu_{\text {SDPF-NR}}$}               &\multicolumn{1}{c}{$\mu_{\text {SDPF-U}}$} 
                         & \multicolumn{1}{c}{$\mu_{\text {SDPF-MU}}$ }\\
\hline\\[-1em]
$^{39}$K & $3/2^+$    & +0.2611\,~\,(3)\,~\,~\,[8]    &+0.3917\,~\,(5)\,[12]    & +0.63  & +0.63  & +0.65  \\[0.1em]
$^{47}$K & $1/2^+$    & +3.8584\,~\,(2)\,[116]    &+1.9292\,~\,(1)\,[58]   & +1.87   & +1.91 & +1.91   \\[0.1em]
$^{49}$K & $1/2^+$     & +2.6772\,(16)\,~\,[80]  & +1.3386\,~\,(8)\,[40] & +1.61  & +1.81 & +1.72 \\[0.1em]
$^{51}$K & $3/2^+$      & +0.3420\,(15)\,~\,[10]    & +0.5129\,(22)\,[15]  & +0.60\, (\rm{E}=0) & +0.63\, (\rm{E}=0) & +0.51\, (\rm{E}=0)   \\[0.1em]
$^{51}$K & ($5/2^+$)      & +0.2279\,(10)\,~\,~\,[7]    &+0.5698\,(25)\,[17]   & +0.84\, (\rm{E}=1725) & +1.40\, (\rm{E}=2018) & +0.75\, (\rm{E}=2264) \\[0.1em]
$^{51}$K & ($7/2^+$)      &  +0.1710\,~\,(8)\,~\,~\,[5]         &+0.5986\,(28)\,[18]               &  -0.13\, (\rm{E}=1793) & -0.04 (\rm{E}=2040)\, & -0.10\, (\rm{E}=2048) \\

\end{tabular} 
\end{ruledtabular}
\end{table*}
The ground-state wave function of the odd-$A$ potassium isotopes is dominated by one proton hole in the $Z = 20$ shell.  The Schmidt moments and thus also the free-nucleon $g$ factors of the relevant single-particle orbits $\pi 1d_{3/2}$, $\pi 2s_{1/2}$ and $\pi 1f_{7/2}$ orbits are very different from each other (respectively +0.08, +5.58 and +1.65).  Therefore their $g$ factors are an excellent probe to monitor in which orbital the unpaired proton occurs.  Because the single nucleon does not appear as a free particle, the experimental values are compared to effective single-nucleon $g$ factors, with  typical values for the $sd$ shell \cite{Richter2008}. The upper part of Fig.~\ref{g-mu} illustrates that the experimental $g$ factors of $^{39-45}$K are all close to the effective value for a hole in the $\pi 1d_{3/2}$ orbit ($g_{\rm{{eff}}} = +0.25$).  Also the $^{51}$K $g$ factors (assuming different spins, given in column 3 of Table \ref{moments}) agree very well with this value.  This confirms the dominant $\pi 1d_{3/2}^{-1}$ component in the ground-state wave function of each of these isotopes, including $^{51}$K. The ground-state spins are known to be $3/2^{+}$ up to $^{45}$K and this result supports the $3/2^{+}$ assignment for the $^{51}$K ground state.  The $^{47}$K $g$ factor is close to the effective single-particle value for a $\pi 2s_{1/2}^{-1}$ hole configuration and its spin/parity is known to be 1/2$^+$ \cite{Touchard1982}.  The ground-state spin of $^{49}$K is established to be also $I=1/2$, but its $g$ factor suggests a rather mixed wave function.

In what follows we focus on further establishing the ground-state spin of $^{51}$K.  For this we compare the experimental magnetic moments to values calculated with different effective shell-model interactions. Calculations have been performed with the SDPF-NR \cite{Retamosa1997,Nummela2001} and its recently upgraded SDPF-U interaction \cite{Nowacki2009} as well as with the recently developed SDPF-MU interaction \cite{UtsunoArXiv} (Table \ref{moments}). The SDPF-NR and SDPF-U interactions have monopole matrix elements that were tuned by fitting experimental spectra of isotopes in this region, from O to Ca. The SPDF-MU interaction is based on the recently developed monopole-based universal interaction $V_{\text{MU}}$ \cite{Otsuka2010} and involves therefore fewer fitted parameters. Calculations are performed with protons restricted to the $sd$ shell (thus only positive parity levels are calculated) and neutrons in the full $pf$ shell. The experimental magnetic moments for the $3/2^+$ ground states of $^{39-45}$K are well reproduced by all effective interactions (lower part of Fig.~\ref{g-mu}), and the agreement for the $3/2^+$ state in $^{51}$K is even better than 0.1 $\mu_N$ for all of them. The lowest positive parity states (5/2$^{+}$, 7/2$^{+}$) appear around 2 MeV and have magnetic moments that deviate significantly more from the experimental values (Table \ref{moments}). Negative parity $3/2^{-}$, $5/2^{-}$, $7/2^{-}$ states, due to a proton excited in the $pf$ shell, all have a magnetic moment that is larger than +3.3 $\mu_{\rm{N}}$, incompatible with the observed small value around +0.5 $\mu_{\rm{N}}$. All arguments together allow to conclude that the ground-state spin/parity of $^{51}$K is $3/2^{+}$.   
  The magnetic moment of the $1/2^+$ ground state in $^{47}$K is also reproduced very well, while the experimental moment of the $I=1/2^+$ ground state of $^{49}$K is somewhat overestimated by all calculations, suggesting that some particular mixing in the wave function is not well taken into account. A simple two-level mixing calculation shows that 25\% mixing of a [$\pi d_{3/2}^{-1}(\nu fp)_{2^{+}}$]$_{{1/2}^{+}}$ allows to reproduce the observed $1/2^{+}$ moment \cite{NeyenProc}.

By establishing the ground-state spins of $^{49}$K and $^{51}$K, we have demonstrated that the gradual reduction of the energy gap between the proton $\pi 2s_{1/2}- \pi 1d_{3/2}$ orbits reaches a minimum around $N=29$ and again increases towards the more neutron rich isotopes.  It is the first time that such a 're-inversion' of single-particle levels is observed and it illustrates how the residual monopole interaction dominates their evolution.

\begin{figure*}
\includegraphics[width=1.0\textwidth]{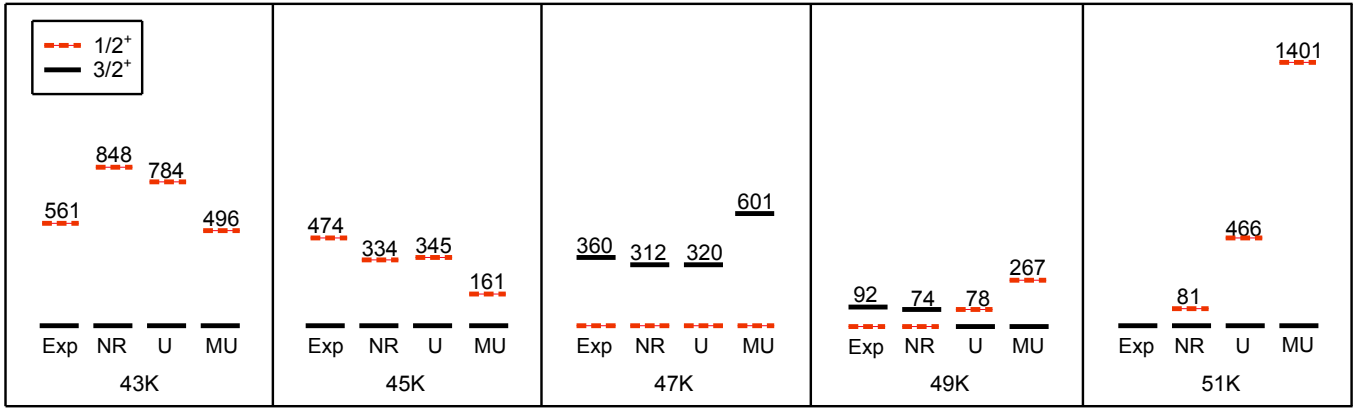}
\caption{\label{energies} (Color online) Experimental energies in odd-K isotopes \cite{Measday2006,Huck1980,Weissman2004,Broda2010,Perrot2006} compared with calculated levels (SDPF-NR,SDPF-U and SDPF-MU interactions; details in text). 3/2$^{+}$ levels are solid black lines and 1/2$^{+}$ dashed red. Ground-state spins for $^{49}$K and $^{51}$K are from this work.}
\end{figure*}

In Fig.~\ref{energies} we compare the experimental $3/2^{+}$ and $1/2^{+}$ levels to those calculated with the different shell-model effective interactions. The SDPF-NR and SDPF-U interactions show the best overall agreement, which is not surprising because their monopole matrix elements were tuned by fitting to experimental spectra, including that of $^{47}$K \cite{Nowacki2009}. With the recently developed SDPF-MU interaction \cite{UtsunoArXiv} a reasonable agreement with the data is found, considering that its cross-shell interaction is described in a functional form using the simple tensor-subtracted monopole evolution as described in \cite{Otsuka2010}, with only six parameters.

In conclusion, the hyperfine structures of atomic $^{49,51}$K isotopes were measured for the first time.  The data establish a ground-state spin $I=1/2$ for $^{49}$K and $I=3/2$ for $^{51}$K.  The magnetic moments $\mu(^{49}\text{K})= +1.3386(8)[40]\, \mu_N$ and $\mu(^{51}\text{K})= +0.5129(22)[15]\, \mu_N$ reveal a mixed configuration for $^{49}$K and a rather pure $\pi 1d_{3/2}^{-1}$ configuration for $^{51}$K. Comparison with shell-model calculations shows good agreement for $^{51}$K, but none of the interactions reproduces the low experimental value of $^{49}$K. Best overall agreement with the ground-state moment and energy levels in $^{49}$K is observed for the SDPF-NR interaction, which predicts the highest mixing with $\pi 1d_{3/2}$ components in its wave function. The experimentally observed evolution of the $1/2^{+}$ and $3/2^{+}$ levels is now established up to $^{51}$K. Different effective interactions predict very different energy gaps between the 3/2$^{+}$ and the first exited 1/2$^+$ level in $^{51}$K.  Along with the current results, spectroscopy of the excited states in $^{51}$K is required to further improve the effective interactions in this region.

\begin{acknowledgments}
This work was supported by the IAP-project P6/23, the FWO-Vlaanderen, NSF grant PHY-1068217, BMBF (05 P12 RDCIC), Max-Planck Society and EU FP7 via ENSAR (no. 262010). We would like to thank to the ISOLDE technical group for their support and assistance.
\end{acknowledgments}

\bibliography{References-G-short}

\begin{thebibliography}{41}%
\makeatletter
\providecommand \@ifxundefined [1]{%
 \@ifx{#1\undefined}
}%
\providecommand \@ifnum [1]{%
 \ifnum #1\expandafter \@firstoftwo
 \else \expandafter \@secondoftwo
 \fi
}%
\providecommand \@ifx [1]{%
 \ifx #1\expandafter \@firstoftwo
 \else \expandafter \@secondoftwo
 \fi
}%
\providecommand \natexlab [1]{#1}%
\providecommand \enquote  [1]{``#1''}%
\providecommand \bibnamefont  [1]{#1}%
\providecommand \bibfnamefont [1]{#1}%
\providecommand \citenamefont [1]{#1}%
\providecommand \href@noop [0]{\@secondoftwo}%
\providecommand \href [0]{\begingroup \@sanitize@url \@href}%
\providecommand \@href[1]{\@@startlink{#1}\@@href}%
\providecommand \@@href[1]{\endgroup#1\@@endlink}%
\providecommand \@sanitize@url [0]{\catcode `\\12\catcode `\$12\catcode
  `\&12\catcode `\#12\catcode `\^12\catcode `\_12\catcode `\%12\relax}%
\providecommand \@@startlink[1]{}%
\providecommand \@@endlink[0]{}%
\providecommand \url  [0]{\begingroup\@sanitize@url \@url }%
\providecommand \@url [1]{\endgroup\@href {#1}{\urlprefix }}%
\providecommand \urlprefix  [0]{URL }%
\providecommand \Eprint [0]{\href }%
\providecommand \doibase [0]{http://dx.doi.org/}%
\providecommand \selectlanguage [0]{\@gobble}%
\providecommand \bibinfo  [0]{\@secondoftwo}%
\providecommand \bibfield  [0]{\@secondoftwo}%
\providecommand \translation [1]{[#1]}%
\providecommand \BibitemOpen [0]{}%
\providecommand \bibitemStop [0]{}%
\providecommand \bibitemNoStop [0]{.\EOS\space}%
\providecommand \EOS [0]{\spacefactor3000\relax}%
\providecommand \BibitemShut  [1]{\csname bibitem#1\endcsname}%
\let\auto@bib@innerbib\@empty
\bibitem [{\citenamefont {Mayer}(1948)}]{Mayer48}%
  \BibitemOpen
  \bibfield  {author} {\bibinfo {author} {\bibfnamefont {M.~G.}\ \bibnamefont
  {Mayer}},\ }\href@noop {} {\bibfield  {journal} {\bibinfo  {journal} {Phys.
  Rev.}\ }\textbf {\bibinfo {volume} {74}},\ \bibinfo {pages} {235} (\bibinfo
  {year} {1948})}\BibitemShut {NoStop}%
\bibitem [{\citenamefont {Mayer}(1949)}]{Mayer49}%
  \BibitemOpen
  \bibfield  {author} {\bibinfo {author} {\bibfnamefont {M.~G.}\ \bibnamefont
  {Mayer}},\ }\href {\doibase 10.1103/PhysRev.75.1969} {\bibfield  {journal}
  {\bibinfo  {journal} {Phys. Rev.}\ }\textbf {\bibinfo {volume} {75}},\
  \bibinfo {pages} {1969} (\bibinfo {year} {1949})}\BibitemShut {NoStop}%
\bibitem [{\citenamefont {Caurier}\ \emph {et~al.}(2005)\citenamefont
  {Caurier}, \citenamefont {Mart\'inez-Pinedo}, \citenamefont {Nowacki},
  \citenamefont {Poves},\ and\ \citenamefont {Zuker}}]{Caurier2005}%
  \BibitemOpen
  \bibfield  {author} {\bibinfo {author} {\bibfnamefont {E.}~\bibnamefont
  {Caurier}}, \bibinfo {author} {\bibfnamefont {G.}~\bibnamefont
  {Mart\'inez-Pinedo}}, \bibinfo {author} {\bibfnamefont {F.}~\bibnamefont
  {Nowacki}}, \bibinfo {author} {\bibfnamefont {A.}~\bibnamefont {Poves}}, \
  and\ \bibinfo {author} {\bibfnamefont {A.~P.}\ \bibnamefont {Zuker}},\ }\href
  {\doibase 10.1103/RevModPhys.77.427} {\bibfield  {journal} {\bibinfo
  {journal} {Rev. Mod. Phys.}\ }\textbf {\bibinfo {volume} {77}},\ \bibinfo
  {pages} {427} (\bibinfo {year} {2005})}\BibitemShut {NoStop}%
\bibitem [{\citenamefont {Neyens}(2011)}]{Neyens2011}%
  \BibitemOpen
  \bibfield  {author} {\bibinfo {author} {\bibfnamefont {G.}~\bibnamefont
  {Neyens}},\ }\href {\doibase 10.1103/PhysRevC.84.064310} {\bibfield
  {journal} {\bibinfo  {journal} {Phys. Rev. C}\ }\textbf {\bibinfo {volume}
  {84}},\ \bibinfo {pages} {064310} (\bibinfo {year} {2011})}\BibitemShut
  {NoStop}%
\bibitem [{\citenamefont {Bastin}\ \emph {et~al.}(2007)\citenamefont {Bastin}
  \emph {et~al.}}]{Bastin2007}%
  \BibitemOpen
  \bibfield  {author} {\bibinfo {author} {\bibfnamefont {B.}~\bibnamefont
  {Bastin}} \emph {et~al.},\ }\href {\doibase 10.1103/PhysRevLett.99.022503}
  {\bibfield  {journal} {\bibinfo  {journal} {Phys. Rev. Lett.}\ }\textbf
  {\bibinfo {volume} {99}},\ \bibinfo {pages} {022503} (\bibinfo {year}
  {2007})}\BibitemShut {NoStop}%
\bibitem [{\citenamefont {Gade}\ \emph {et~al.}(2006)\citenamefont {Gade} \emph
  {et~al.}}]{Gade2006}%
  \BibitemOpen
  \bibfield  {author} {\bibinfo {author} {\bibfnamefont {A.}~\bibnamefont
  {Gade}} \emph {et~al.},\ }\href {\doibase 10.1103/PhysRevC.74.034322}
  {\bibfield  {journal} {\bibinfo  {journal} {Phys. Rev. C}\ }\textbf {\bibinfo
  {volume} {74}},\ \bibinfo {pages} {034322} (\bibinfo {year}
  {2006})}\BibitemShut {NoStop}%
\bibitem [{\citenamefont {Flanagan}\ \emph {et~al.}(2009)\citenamefont
  {Flanagan} \emph {et~al.}}]{Flanagan2009}%
  \BibitemOpen
  \bibfield  {author} {\bibinfo {author} {\bibfnamefont {K.~T.}\ \bibnamefont
  {Flanagan}} \emph {et~al.},\ }\href {\doibase 10.1103/PhysRevLett.103.142501}
  {\bibfield  {journal} {\bibinfo  {journal} {Phys. Rev. Lett.}\ }\textbf
  {\bibinfo {volume} {103}},\ \bibinfo {pages} {142501} (\bibinfo {year}
  {2009})}\BibitemShut {NoStop}%
\bibitem [{\citenamefont {Otsuka}\ \emph {et~al.}(2001)\citenamefont {Otsuka}
  \emph {et~al.}}]{Otsuka2001}%
  \BibitemOpen
  \bibfield  {author} {\bibinfo {author} {\bibfnamefont {T.}~\bibnamefont
  {Otsuka}} \emph {et~al.},\ }\href@noop {} {\bibfield  {journal} {\bibinfo
  {journal} {Phys. Rev. Lett.}\ }\textbf {\bibinfo {volume} {87}},\ \bibinfo
  {pages} {082502} (\bibinfo {year} {2001})}\BibitemShut {NoStop}%
\bibitem [{\citenamefont {Otsuka}\ \emph {et~al.}(2005)\citenamefont {Otsuka},
  \citenamefont {Suzuki}, \citenamefont {Fujimoto}, \citenamefont {Grawe},\
  and\ \citenamefont {Akaishi}}]{Otsuka2005}%
  \BibitemOpen
  \bibfield  {author} {\bibinfo {author} {\bibfnamefont {T.}~\bibnamefont
  {Otsuka}}, \bibinfo {author} {\bibfnamefont {T.}~\bibnamefont {Suzuki}},
  \bibinfo {author} {\bibfnamefont {R.}~\bibnamefont {Fujimoto}}, \bibinfo
  {author} {\bibfnamefont {H.}~\bibnamefont {Grawe}}, \ and\ \bibinfo {author}
  {\bibfnamefont {Y.}~\bibnamefont {Akaishi}},\ }\href {\doibase
  10.1103/PhysRevLett.95.232502} {\bibfield  {journal} {\bibinfo  {journal}
  {Phys. Rev. Lett.}\ }\textbf {\bibinfo {volume} {95}},\ \bibinfo {pages}
  {232502} (\bibinfo {year} {2005})}\BibitemShut {NoStop}%
\bibitem [{\citenamefont {Otsuka}\ \emph {et~al.}(2006)\citenamefont {Otsuka},
  \citenamefont {Matsuo},\ and\ \citenamefont {Abe}}]{Otsuka2006}%
  \BibitemOpen
  \bibfield  {author} {\bibinfo {author} {\bibfnamefont {T.}~\bibnamefont
  {Otsuka}}, \bibinfo {author} {\bibfnamefont {T.}~\bibnamefont {Matsuo}}, \
  and\ \bibinfo {author} {\bibfnamefont {D.}~\bibnamefont {Abe}},\ }\href
  {\doibase 10.1103/PhysRevLett.97.162501} {\bibfield  {journal} {\bibinfo
  {journal} {Phys. Rev. Lett.}\ }\textbf {\bibinfo {volume} {97}},\ \bibinfo
  {pages} {162501} (\bibinfo {year} {2006})}\BibitemShut {NoStop}%
\bibitem [{\citenamefont {Smirnova}\ \emph {et~al.}(2010)\citenamefont
  {Smirnova}, \citenamefont {Bally}, \citenamefont {Heyde}, \citenamefont
  {Nowacki},\ and\ \citenamefont {Sieja}}]{Smirnova2010}%
  \BibitemOpen
  \bibfield  {author} {\bibinfo {author} {\bibfnamefont {N.~A.}\ \bibnamefont
  {Smirnova}}, \bibinfo {author} {\bibfnamefont {B.}~\bibnamefont {Bally}},
  \bibinfo {author} {\bibfnamefont {K.}~\bibnamefont {Heyde}}, \bibinfo
  {author} {\bibfnamefont {F.}~\bibnamefont {Nowacki}}, \ and\ \bibinfo
  {author} {\bibfnamefont {K.}~\bibnamefont {Sieja}},\ }\href {\doibase
  doi:10.1016/j.physletb.2010.02.051} {\bibfield  {journal} {\bibinfo
  {journal} {Phys. Lett. B}\ }\textbf {\bibinfo {volume} {686}},\ \bibinfo
  {pages} {109} (\bibinfo {year} {2010})}\BibitemShut {NoStop}%
\bibitem [{\citenamefont {Otsuka}\ \emph {et~al.}(2010)\citenamefont {Otsuka}
  \emph {et~al.}}]{Otsuka2010}%
  \BibitemOpen
  \bibfield  {author} {\bibinfo {author} {\bibfnamefont {T.}~\bibnamefont
  {Otsuka}} \emph {et~al.},\ }\href@noop {} {\bibfield  {journal} {\bibinfo
  {journal} {Phys. Rev. Lett.}\ }\textbf {\bibinfo {volume} {104}},\ \bibinfo
  {pages} {012501} (\bibinfo {year} {2010})}\BibitemShut {NoStop}%
\bibitem [{\citenamefont {Grasso}\ \emph {et~al.}(2007)\citenamefont {Grasso},
  \citenamefont {Ma}, \citenamefont {Khan}, \citenamefont {Margueron},\ and\
  \citenamefont {VanGiai}}]{Grasso2007}%
  \BibitemOpen
  \bibfield  {author} {\bibinfo {author} {\bibfnamefont {M.}~\bibnamefont
  {Grasso}}, \bibinfo {author} {\bibfnamefont {Z.~Y.}\ \bibnamefont {Ma}},
  \bibinfo {author} {\bibfnamefont {E.}~\bibnamefont {Khan}}, \bibinfo {author}
  {\bibfnamefont {J.}~\bibnamefont {Margueron}}, \ and\ \bibinfo {author}
  {\bibfnamefont {N.}~\bibnamefont {VanGiai}},\ }\href {\doibase
  10.1103/PhysRevC.76.044319} {\bibfield  {journal} {\bibinfo  {journal} {Phys.
  Rev. C}\ }\textbf {\bibinfo {volume} {76}},\ \bibinfo {pages} {044319}
  (\bibinfo {year} {2007})}\BibitemShut {NoStop}%
\bibitem [{\citenamefont {Nowacki}\ and\ \citenamefont
  {Poves}(2009)}]{Nowacki2009}%
  \BibitemOpen
  \bibfield  {author} {\bibinfo {author} {\bibfnamefont {F.}~\bibnamefont
  {Nowacki}}\ and\ \bibinfo {author} {\bibfnamefont {A.}~\bibnamefont
  {Poves}},\ }\href {\doibase 10.1103/PhysRevC.79.014310} {\bibfield  {journal}
  {\bibinfo  {journal} {Phys. Rev. C}\ }\textbf {\bibinfo {volume} {79}},\
  \bibinfo {pages} {014310} (\bibinfo {year} {2009})}\BibitemShut {NoStop}%
\bibitem [{\citenamefont {Zalewski}\ \emph {et~al.}(2009)\citenamefont
  {Zalewski} \emph {et~al.}}]{Zalewski2009}%
  \BibitemOpen
  \bibfield  {author} {\bibinfo {author} {\bibfnamefont {M.}~\bibnamefont
  {Zalewski}} \emph {et~al.},\ }\href {\doibase 10.1103/PhysRevC.80.064307}
  {\bibfield  {journal} {\bibinfo  {journal} {Phys. Rev. C}\ }\textbf {\bibinfo
  {volume} {80}},\ \bibinfo {pages} {064307} (\bibinfo {year}
  {2009})}\BibitemShut {NoStop}%
\bibitem [{\citenamefont {Kaneko}\ \emph {et~al.}(2011)\citenamefont {Kaneko},
  \citenamefont {Sun}, \citenamefont {Mizusaki},\ and\ \citenamefont
  {Hasegawa}}]{Kaneko2011}%
  \BibitemOpen
  \bibfield  {author} {\bibinfo {author} {\bibfnamefont {K.}~\bibnamefont
  {Kaneko}}, \bibinfo {author} {\bibfnamefont {Y.}~\bibnamefont {Sun}},
  \bibinfo {author} {\bibfnamefont {T.}~\bibnamefont {Mizusaki}}, \ and\
  \bibinfo {author} {\bibfnamefont {M.}~\bibnamefont {Hasegawa}},\ }\href
  {\doibase 10.1103/PhysRevC.83.014320} {\bibfield  {journal} {\bibinfo
  {journal} {Phys. Rev. C}\ }\textbf {\bibinfo {volume} {83}},\ \bibinfo
  {pages} {014320} (\bibinfo {year} {2011})}\BibitemShut {NoStop}%
\bibitem [{\citenamefont {Y.Z.Wang}\ \emph {et~al.}(2011)\citenamefont
  {Y.Z.Wang}, \citenamefont {J.Z.Gu}, \citenamefont {X.Z.Zhang},\ and\
  \citenamefont {J.M.Dong}}]{Wang2011}%
  \BibitemOpen
  \bibfield  {author} {\bibinfo {author} {\bibnamefont {Y.Z.Wang}}, \bibinfo
  {author} {\bibnamefont {J.Z.Gu}}, \bibinfo {author} {\bibnamefont
  {X.Z.Zhang}}, \ and\ \bibinfo {author} {\bibnamefont {J.M.Dong}},\ }\href
  {\doibase 10.1103/PhysRevC.84.044333} {\bibfield  {journal} {\bibinfo
  {journal} {Phys. Rev. C}\ }\textbf {\bibinfo {volume} {84}},\ \bibinfo
  {pages} {044333} (\bibinfo {year} {2011})}\BibitemShut {NoStop}%
\bibitem [{\citenamefont {Touchard}\ \emph {et~al.}(1982)\citenamefont
  {Touchard} \emph {et~al.}}]{Touchard1982}%
  \BibitemOpen
  \bibfield  {author} {\bibinfo {author} {\bibfnamefont {F.}~\bibnamefont
  {Touchard}} \emph {et~al.},\ }\href {\doibase 10.1016/j.ppnp.2008.05.001}
  {\bibfield  {journal} {\bibinfo  {journal} {Phys. Lett. B}\ }\textbf
  {\bibinfo {volume} {108}},\ \bibinfo {pages} {169} (\bibinfo {year}
  {1982})}\BibitemShut {NoStop}%
\bibitem [{\citenamefont {Ollier}\ \emph {et~al.}(2003)\citenamefont {Ollier}
  \emph {et~al.}}]{Ollier2003}%
  \BibitemOpen
  \bibfield  {author} {\bibinfo {author} {\bibfnamefont {J.}~\bibnamefont
  {Ollier}} \emph {et~al.},\ }\href {\doibase 10.1103/PhysRevC.67.024302}
  {\bibfield  {journal} {\bibinfo  {journal} {Phys. Rev. C}\ }\textbf {\bibinfo
  {volume} {67}},\ \bibinfo {pages} {024302} (\bibinfo {year}
  {2003})}\BibitemShut {NoStop}%
\bibitem [{\citenamefont {Broda}\ \emph {et~al.}(2010)\citenamefont {Broda}
  \emph {et~al.}}]{Broda2010}%
  \BibitemOpen
  \bibfield  {author} {\bibinfo {author} {\bibfnamefont {R.}~\bibnamefont
  {Broda}} \emph {et~al.},\ }\href {\doibase 10.1103/PhysRevC.82.034319}
  {\bibfield  {journal} {\bibinfo  {journal} {Phys. Rev. C}\ }\textbf {\bibinfo
  {volume} {82}},\ \bibinfo {pages} {034319} (\bibinfo {year}
  {2010})}\BibitemShut {NoStop}%
\bibitem [{\citenamefont {Carraz}\ \emph {et~al.}(1982)\citenamefont {Carraz}
  \emph {et~al.}}]{Carraz1982}%
  \BibitemOpen
  \bibfield  {author} {\bibinfo {author} {\bibfnamefont {L.~C.}\ \bibnamefont
  {Carraz}} \emph {et~al.},\ }\href {\doibase 10.1016/0370-2693(82)91104-2}
  {\bibfield  {journal} {\bibinfo  {journal} {Physics Letters B}\ }\textbf
  {\bibinfo {volume} {109}},\ \bibinfo {pages} {419 } (\bibinfo {year}
  {1982})}\BibitemShut {NoStop}%
\bibitem [{\citenamefont {Perrot}\ \emph {et~al.}(2006)\citenamefont {Perrot}
  \emph {et~al.}}]{Perrot2006}%
  \BibitemOpen
  \bibfield  {author} {\bibinfo {author} {\bibfnamefont {F.}~\bibnamefont
  {Perrot}} \emph {et~al.},\ }\href {\doibase 10.1103/PhysRevC.74.014313}
  {\bibfield  {journal} {\bibinfo  {journal} {Phys. Rev. C}\ }\textbf {\bibinfo
  {volume} {74}},\ \bibinfo {pages} {014313} (\bibinfo {year}
  {2006})}\BibitemShut {NoStop}%
\bibitem [{\citenamefont {Franberg}\ \emph {et~al.}(2008)\citenamefont
  {Franberg} \emph {et~al.}}]{Franberg2008}%
  \BibitemOpen
  \bibfield  {author} {\bibinfo {author} {\bibfnamefont {H.}~\bibnamefont
  {Franberg}} \emph {et~al.},\ }\href@noop {} {\bibfield  {journal} {\bibinfo
  {journal} {Nucl. Inst. and Meth. in Phys. Res. B}\ }\textbf {\bibinfo
  {volume} {266}},\ \bibinfo {pages} {4502} (\bibinfo {year}
  {2008})}\BibitemShut {NoStop}%
\bibitem [{\citenamefont {Mueller}\ \emph {et~al.}(1983)\citenamefont {Mueller}
  \emph {et~al.}}]{Mueller1983}%
  \BibitemOpen
  \bibfield  {author} {\bibinfo {author} {\bibfnamefont {A.~C.}\ \bibnamefont
  {Mueller}} \emph {et~al.},\ }\href@noop {} {\bibfield  {journal} {\bibinfo
  {journal} {Nucl. Phys. A}\ }\textbf {\bibinfo {volume} {403}},\ \bibinfo
  {pages} {234} (\bibinfo {year} {1983})}\BibitemShut {NoStop}%
\bibitem [{\citenamefont {Kreim}\ \emph {et~al.}()\citenamefont {Kreim} \emph
  {et~al.}}]{Kreim2013}%
  \BibitemOpen
  \bibfield  {author} {\bibinfo {author} {\bibfnamefont {K.}~\bibnamefont
  {Kreim}} \emph {et~al.},\ }\href@noop {} {}\bibinfo {note} {In
  preparation}\BibitemShut {NoStop}%
\bibitem [{\citenamefont {Wang}\ \emph {et~al.}(2012)\citenamefont {Wang} \emph
  {et~al.}}]{Wang2012}%
  \BibitemOpen
  \bibfield  {author} {\bibinfo {author} {\bibfnamefont {M.}~\bibnamefont
  {Wang}} \emph {et~al.},\ }\href@noop {} {\bibfield  {journal} {\bibinfo
  {journal} {Chinese Physics C}\ }\textbf {\bibinfo {volume} {36}},\ \bibinfo
  {pages} {1603�2014} (\bibinfo {year} {2012})}\BibitemShut {NoStop}%
\bibitem [{\citenamefont {Lapierre}\ \emph {et~al.}(2012)\citenamefont
  {Lapierre} \emph {et~al.}}]{Lapierre2012}%
  \BibitemOpen
  \bibfield  {author} {\bibinfo {author} {\bibfnamefont {A.}~\bibnamefont
  {Lapierre}} \emph {et~al.},\ }\href {\doibase 10.1103/PhysRevC.85.024317}
  {\bibfield  {journal} {\bibinfo  {journal} {Phys. Rev. C}\ }\textbf {\bibinfo
  {volume} {85}},\ \bibinfo {pages} {024317} (\bibinfo {year}
  {2012})}\BibitemShut {NoStop}%
\bibitem [{\citenamefont {Gallant}\ \emph {et~al.}(2012)\citenamefont {Gallant}
  \emph {et~al.}}]{Gallant2012}%
  \BibitemOpen
  \bibfield  {author} {\bibinfo {author} {\bibfnamefont {A.~T.}\ \bibnamefont
  {Gallant}} \emph {et~al.},\ }\href {\doibase 10.1103/PhysRevLett.109.032506}
  {\bibfield  {journal} {\bibinfo  {journal} {Phys. Rev. Lett.}\ }\textbf
  {\bibinfo {volume} {109}},\ \bibinfo {pages} {032506} (\bibinfo {year}
  {2012})}\BibitemShut {NoStop}%
\bibitem [{\citenamefont {Flanagan}\ \emph {et~al.}(2010)\citenamefont
  {Flanagan} \emph {et~al.}}]{Flanagan2010}%
  \BibitemOpen
  \bibfield  {author} {\bibinfo {author} {\bibfnamefont {K.~T.}\ \bibnamefont
  {Flanagan}} \emph {et~al.},\ }\href {\doibase 10.1103/PhysRevC.82.041302}
  {\bibfield  {journal} {\bibinfo  {journal} {Phys. Rev. C}\ }\textbf {\bibinfo
  {volume} {82}},\ \bibinfo {pages} {041302(R)} (\bibinfo {year}
  {2010})}\BibitemShut {NoStop}%
\bibitem [{\citenamefont {Beckmann}\ \emph {et~al.}(1974)\citenamefont
  {Beckmann}, \citenamefont {B\"{o}klen},\ and\ \citenamefont
  {Elke}}]{Beckmann1974}%
  \BibitemOpen
  \bibfield  {author} {\bibinfo {author} {\bibfnamefont {A.}~\bibnamefont
  {Beckmann}}, \bibinfo {author} {\bibfnamefont {K.~D.}\ \bibnamefont
  {B\"{o}klen}}, \ and\ \bibinfo {author} {\bibfnamefont {D.}~\bibnamefont
  {Elke}},\ }\href {http://dx.doi.org/10.1007/BF01680407} {\bibfield  {journal}
  {\bibinfo  {journal} {Zeitschrift f\"{u}r Physik A Hadrons and Nuclei}\
  }\textbf {\bibinfo {volume} {270}},\ \bibinfo {pages} {173} (\bibinfo {year}
  {1974})}\BibitemShut {NoStop}%
\bibitem [{\citenamefont {Eisinger}\ \emph {et~al.}(1952)\citenamefont
  {Eisinger}, \citenamefont {Bederson},\ and\ \citenamefont
  {Feld}}]{Eisinger1952}%
  \BibitemOpen
  \bibfield  {author} {\bibinfo {author} {\bibfnamefont {J.~T.}\ \bibnamefont
  {Eisinger}}, \bibinfo {author} {\bibfnamefont {B.}~\bibnamefont {Bederson}},
  \ and\ \bibinfo {author} {\bibfnamefont {B.~T.}\ \bibnamefont {Feld}},\
  }\href@noop {} {\bibfield  {journal} {\bibinfo  {journal} {Phys. Rev.}\
  }\textbf {\bibinfo {volume} {86}},\ \bibinfo {pages} {73} (\bibinfo {year}
  {1952})}\BibitemShut {NoStop}%
\bibitem [{\citenamefont {Chan}\ \emph {et~al.}(1969)\citenamefont {Chan},
  \citenamefont {Cohen},\ and\ \citenamefont {Silsbee}}]{Chan1969}%
  \BibitemOpen
  \bibfield  {author} {\bibinfo {author} {\bibfnamefont {Y.~W.}\ \bibnamefont
  {Chan}}, \bibinfo {author} {\bibfnamefont {V.~W.}\ \bibnamefont {Cohen}}, \
  and\ \bibinfo {author} {\bibfnamefont {H.~B.}\ \bibnamefont {Silsbee}},\
  }\href@noop {} {\bibfield  {journal} {\bibinfo  {journal} {Phys. Rev.}\
  }\textbf {\bibinfo {volume} {184}},\ \bibinfo {pages} {1102} (\bibinfo {year}
  {1969})}\BibitemShut {NoStop}%
\bibitem [{\citenamefont {Bohr}(1951)}]{Bohr1951}%
  \BibitemOpen
  \bibfield  {author} {\bibinfo {author} {\bibfnamefont {A.}~\bibnamefont
  {Bohr}},\ }\href@noop {} {\bibfield  {journal} {\bibinfo  {journal} {Phys.
  Rev.}\ }\textbf {\bibinfo {volume} {81}},\ \bibinfo {pages} {331} (\bibinfo
  {year} {1951})}\BibitemShut {NoStop}%
\bibitem [{\citenamefont {Richter}\ \emph {et~al.}(2008)\citenamefont
  {Richter}, \citenamefont {Mkhize},\ and\ \citenamefont
  {Brown}}]{Richter2008}%
  \BibitemOpen
  \bibfield  {author} {\bibinfo {author} {\bibfnamefont {W.~A.}\ \bibnamefont
  {Richter}}, \bibinfo {author} {\bibfnamefont {S.}~\bibnamefont {Mkhize}}, \
  and\ \bibinfo {author} {\bibfnamefont {B.~A.}\ \bibnamefont {Brown}},\
  }\href@noop {} {\bibfield  {journal} {\bibinfo  {journal} {Phys. Rev. C}\
  }\textbf {\bibinfo {volume} {78}},\ \bibinfo {pages} {064302} (\bibinfo
  {year} {2008})}\BibitemShut {NoStop}%
\bibitem [{\citenamefont {Retamosa}\ \emph {et~al.}(1997)\citenamefont
  {Retamosa}, \citenamefont {Caurier}, \citenamefont {Nowacki},\ and\
  \citenamefont {Poves}}]{Retamosa1997}%
  \BibitemOpen
  \bibfield  {author} {\bibinfo {author} {\bibfnamefont {J.}~\bibnamefont
  {Retamosa}}, \bibinfo {author} {\bibfnamefont {E.}~\bibnamefont {Caurier}},
  \bibinfo {author} {\bibfnamefont {F.}~\bibnamefont {Nowacki}}, \ and\
  \bibinfo {author} {\bibfnamefont {A.}~\bibnamefont {Poves}},\ }\href
  {\doibase 10.1103/PhysRevC.55.1266} {\bibfield  {journal} {\bibinfo
  {journal} {Phys. Rev. C}\ }\textbf {\bibinfo {volume} {55}},\ \bibinfo
  {pages} {1266} (\bibinfo {year} {1997})}\BibitemShut {NoStop}%
\bibitem [{\citenamefont {Nummela}\ \emph {et~al.}(2001)\citenamefont {Nummela}
  \emph {et~al.}}]{Nummela2001}%
  \BibitemOpen
  \bibfield  {author} {\bibinfo {author} {\bibfnamefont {S.}~\bibnamefont
  {Nummela}} \emph {et~al.},\ }\href {\doibase 10.1103/PhysRevC.64.054313}
  {\bibfield  {journal} {\bibinfo  {journal} {Phys. Rev. C}\ }\textbf {\bibinfo
  {volume} {64}},\ \bibinfo {pages} {054313} (\bibinfo {year}
  {2001})}\BibitemShut {NoStop}%
\bibitem [{\citenamefont {Utsuno}\ \emph {et~al.}(2012)\citenamefont {Utsuno}
  \emph {et~al.}}]{UtsunoArXiv}%
  \BibitemOpen
  \bibfield  {author} {\bibinfo {author} {\bibfnamefont {Y.}~\bibnamefont
  {Utsuno}} \emph {et~al.},\ }\href@noop {} {\bibfield  {journal} {\bibinfo
  {journal} {Phys. Rev. C}\ }\textbf {\bibinfo {volume} {86}},\ \bibinfo
  {pages} {051301(R)} (\bibinfo {year} {2012})}\BibitemShut {NoStop}%
\bibitem [{\citenamefont {Neyens}()}]{NeyenProc}%
  \BibitemOpen
  \bibfield  {author} {\bibinfo {author} {\bibfnamefont {G.}~\bibnamefont
  {Neyens}},\ }\href@noop {} {}\bibinfo {note} {Proc. Int. Sym. "Exotic Nuclear
  Structure From Nucleons" (ENSFN 2012), Tokyo, Japan, Oct.2012 to be published
  in Journal of Physics: Conference Series (2013)}\BibitemShut {NoStop}%
\bibitem [{\citenamefont {Measday}\ and\ \citenamefont
  {Stocki}(2006)}]{Measday2006}%
  \BibitemOpen
  \bibfield  {author} {\bibinfo {author} {\bibfnamefont {D.~F.}\ \bibnamefont
  {Measday}}\ and\ \bibinfo {author} {\bibfnamefont {T.~J.}\ \bibnamefont
  {Stocki}},\ }\href@noop {} {\bibfield  {journal} {\bibinfo  {journal} {Phys.
  Rev. C}\ }\textbf {\bibinfo {volume} {73}},\ \bibinfo {pages} {045501}
  (\bibinfo {year} {2006})}\BibitemShut {NoStop}%
\bibitem [{\citenamefont {Huck}\ \emph {et~al.}(1980)\citenamefont {Huck} \emph
  {et~al.}}]{Huck1980}%
  \BibitemOpen
  \bibfield  {author} {\bibinfo {author} {\bibfnamefont {A.}~\bibnamefont
  {Huck}} \emph {et~al.},\ }\href@noop {} {\bibfield  {journal} {\bibinfo
  {journal} {Phys. Rev. C}\ }\textbf {\bibinfo {volume} {21}},\ \bibinfo
  {pages} {712} (\bibinfo {year} {1980})}\BibitemShut {NoStop}%
\bibitem [{\citenamefont {Weissman}\ \emph {et~al.}(2004)\citenamefont
  {Weissman} \emph {et~al.}}]{Weissman2004}%
  \BibitemOpen
  \bibfield  {author} {\bibinfo {author} {\bibfnamefont {L.}~\bibnamefont
  {Weissman}} \emph {et~al.},\ }\href@noop {} {\bibfield  {journal} {\bibinfo
  {journal} {Phys. Rev. C}\ }\textbf {\bibinfo {volume} {70}},\ \bibinfo
  {pages} {024304} (\bibinfo {year} {2004})}\BibitemShut {NoStop}%
\end{thebibliography}%

\end{document}